\documentclass[
 reprint,
superscriptaddress,
 amsmath,amssymb,
 aps,
 prl,
]{revtex4-2}

\usepackage[T1]{fontenc}

\usepackage{graphicx}
\usepackage{dcolumn}
\usepackage{bm}
\usepackage{hyperref}
\usepackage[mathlines]{lineno}
\usepackage{braket}

\hypersetup{
    colorlinks=true,
    linkcolor=blue,
    citecolor=blue,
    urlcolor=blue
}
\usepackage{cleveref}
\crefname{equation}{Eq.}{Eqs.}
\Crefname{equation}{Eq.}{Eqs.}

\crefname{figure}{Fig.}{Figs.}
\Crefname{figure}{Fig.}{Figs.}

\newcommand{\subfigrefbase}[3]{#1~\hyperref[#2]{\ref*{#2}#3}}
\newcommand{\subfigref}[2]{\subfigrefbase{Fig.}{#1}{#2}}
\newcommand{\subfigsref}[2]{\subfigrefbase{Figs.}{#1}{#2}}
\newcommand{\subfignodotref}[2]{\subfigrefbase{Figure}{#1}{#2}}
\newcommand{\equationnodotref}[1]{Equation~(\hyperref[#1]{\ref*{#1}})}

\newcommand{\subfigpairref}[3]{%
Figs.~%
\hyperref[#1]{\ref*{#1}#2} and %
\hyperref[#1]{\ref*{#1}#3}%
}
\usepackage{orcidlink}
\begin{document}

\preprint{APS/123-QED}

\title{Barnett effect in rotating spinor dipolar quantum droplets}

\author{Donghao Yan}
\affiliation{Department of Engineering Science, University of Electro-Communications, Tokyo 182-8585, Japan
}
\author{Shaoxiong Li}
\affiliation{Computational Materials Science Research Team, RIKEN Center for Computational Science (R-CCS), Kobe, Hyogo, Japan
}
\author{Hiroki Saito}
\affiliation{Department of Engineering Science, University of Electro-Communications, Tokyo 182-8585, Japan
}

\date{\today}

\begin{abstract}
We propose releasing the spin degree of freedom to stabilize the
vortex state in self-bound droplets of dipolar Bose-Einstein condensates. 
When a vortex is embedded into the droplet, spontaneous magnetization
arises in the axial direction via a mechanism similar to the Barnett
effect; that is, the orbital angular momentum is transferred to the spin angular momentum.
When an external magnetic field is applied to the spontaneously
magnetized droplet, the entire atomic cloud starts to rotate without
changing its shape, which can be regarded as mechanical Larmor
precession of a macroscopic object.
A chirally different pair of droplets can form a stable bound state
because of the attractive interaction between the spontaneously magnetized
droplets.
\end{abstract}

\maketitle

In the past decade, self-bound states of Bose-Einstein condensates
(BECs), referred to as quantum droplets~\cite{Petrov}, have
attracted much interest.
In quantum droplets, the balance between the attractive mean-field
interaction and the repulsive beyond-mean-field effect~\cite{LHY}
prevents the system from expansion and collapse in free space.
Quantum droplets were experimentally realized in BECs with
a magnetic dipole-dipole interaction (DDI)~\cite{Kadau,
  8PhysRevA.96.053630, 4schmitt2016self, 5PhysRevLett.116.215301,
  6PhysRevX.6.041039, 7PhysRevLett.120.160402} and in binary mixtures
of BECs~\cite{Cabrera, Semeghini, Cheiney}; a wide variety of
theoretical studies have been performed on these
systems~\cite{waechtler2016,waechtler2016b,Saito,Bisset,macia2016droplets,Baillie2017,lima2011,lima2012,15PhysRevResearch.1.033088,oldziejewski2020strongly,30bisset2021,smith2021quantum,Boninsegni,20PhysRevLett.130.043401}. 

Such a novel state of BECs naturally raises the question of whether
self-bound droplets can host topological excitations,
such as quantized vortices. 
Because of the centrifugal force and attractive interaction, self-bound
droplets with quantized vortices are prone to azimuthal
instability~\cite{saito2002, Hung2025}, which splits the droplets into fragments~\cite{cidrim2018vortices}.
An external trap potential can effectively suppress this instability
and stabilize quantum droplets with vortices~\cite{qvd-bessel, 17PhysRevLett.123.133901,vqd-harmonic,vqd-strongtrap,vqd-annulartrap,fast-qd,Liu2023}.
In free space without a trap potential, self-bound
droplets of binary BECs have been shown to stably retain quantized vortices~\cite{2d-vqd, stability_vqd,vqd-2cpn,3d-vqd-ss}. 
Droplets of binary BECs rotating in trap potentials have also been studied~\cite{Reimann2019, Ancilotto2022, Kavoulakis2020, Kartashov2021, Ogren2024}.
Stable vortex droplets can also exist in a dipolar BEC, in which the
atomic spins are polarized in the same direction by an external magnetic
field~\cite{Bao2018,ani-vqd, LZhang2024}.
By contrast, in this Letter, we propose dipolar BECs with spin
degrees of freedom -- spinor dipolar BECs~\cite{Pasquiou2011, Matsu2026,kawaguchi2010,hoshi2010, Swislocki,26PhysRevLett.98.110406,  kudo2010hydrodynamic, Oshima, lepoutre2018spin, Yi, Kawaguchi2,Takahashi,Huhtamaki, Simula, liao2017anisotropic, Borgh} -- for producing stable
vortex quantum droplets. Spinor dipolar BECs have been realized experimentally by shielding from external magnetic fields \cite{Pasquiou2011, lepoutre2018spin,Matsu2026}.

\begin{figure} 
    \centering
    \includegraphics[width=\linewidth]{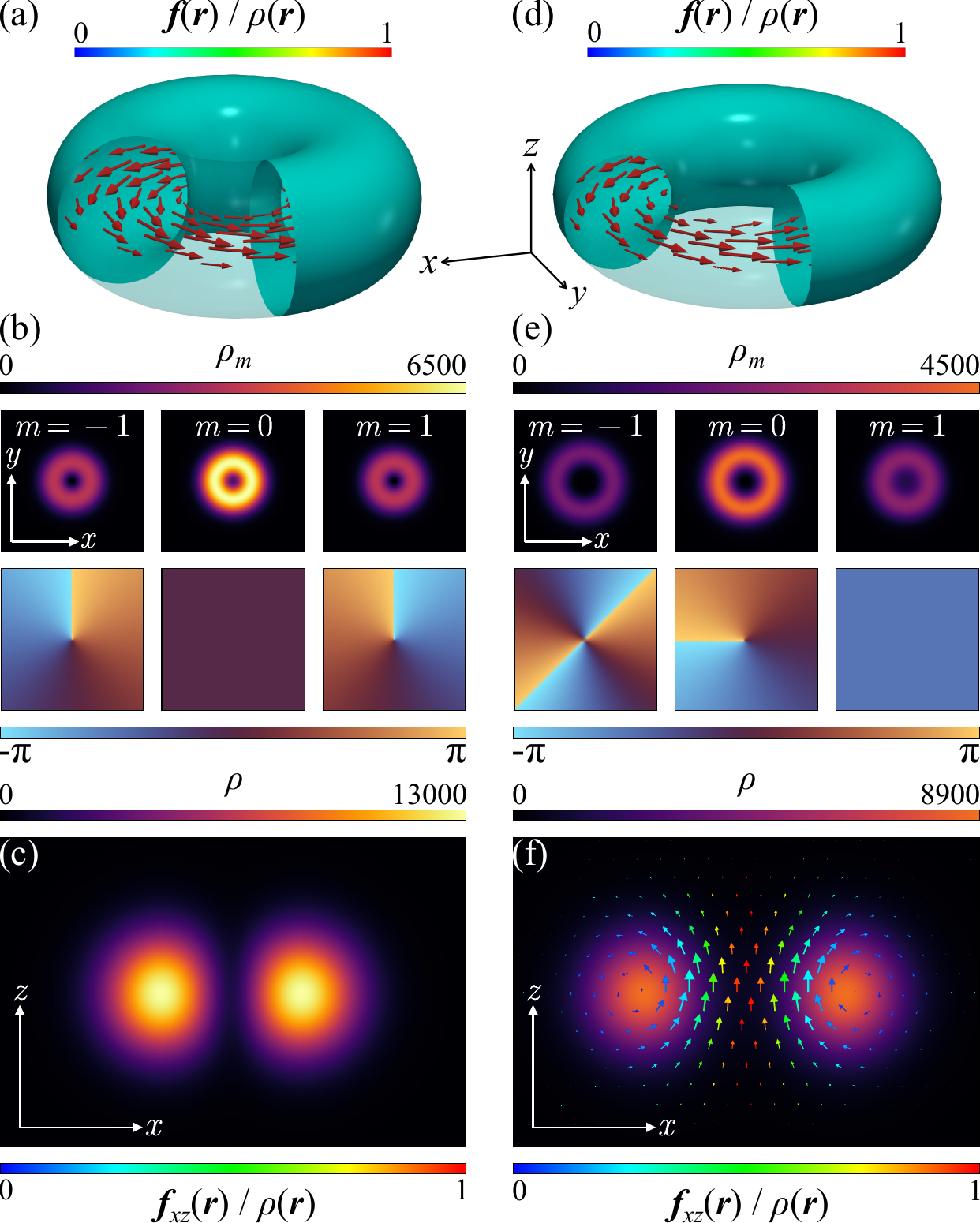}
    \caption{(a-c) Nonrotating ground state and (d-f) rotating ($\ell=1$) stationary state of self-bound droplets for $F=1$, $N=15000$, $\varepsilon_{\rm dd}=1.2$, and $\bm B=0$. (a, d) Isodensity surfaces at half the maximum density. Arrows represent the spin vectors $\bm f(\bm r)$ on the $z=0$ plane, and their color shows the polarization $\bm f \left(\bm r\right)/\rho\left(\bm r\right)$. (b, e) Density (upper panels) and phase (lower panels) distributions of component $m$ on the $z=0$ plane. The size of each panel is $0.07\times0.07$. (c, f) Total density distributions on the $y=0$ plane. Arrows represent the projected spin vectors $\bm{f}_{xz}(\bm{r}) = (f_x(\bm{r}), f_z(\bm{r}))$ and their color shows $|\bm{f}_{xz}(\bm{r})| / \rho(\bm{r})$. The size of each panel is $0.12\times0.08$. The system has translational and rotational symmetry; for convenience, the axis of the torus is taken to be the $z$-axis with its center at the origin.}
    \label{fig:panorama}
\end{figure}

In a spinor dipolar BEC, the spatial distribution of spin vectors
is determined so as to reduce the magnetostatic energy~\cite{Yi, Kawaguchi2, Takahashi}. 
For a self-bound droplet of a spinor dipolar BEC, the magnetic
flux-closure structure, in which the density distribution has a torus
shape and the spin vectors circulate along the torus
[\subfigref{fig:panorama}{(a)}], is the most efficient spin 
distribution to minimize the magnetostatic energy~\cite{Li2024}.
This torus-shaped density distribution is convenient for stabilizing a
vortex because the vortex core is pinned at the torus hole.
In fact, we will show that the vortex embedded in the torus-shaped
spinor dipolar droplet is robust against external disturbances and
that no azimuthal instability arises in the system.

An interesting property of the rotating spinor dipolar droplet is
that it exhibits net magnetization in the axial direction of the torus.
The mechanism responsible for this spontaneous magnetization resembles the Barnett
effect~\cite{Barnett}; that is, the orbital angular momentum is
transferred to the spin angular momentum, resulting in the
rotation-induced magnetization in an uncharged system.
We demonstrate two intriguing phenomena arising from this
spontaneous magnetization of self-bound droplets.
The first phenomenon is macroscopic mechanical Larmor precession:
when an external magnetic field is applied, the total magnetization
of the droplet undergoes Larmor precession, which causes mechanical
rotation of the entire droplet.
The second phenomenon is the formation of a bound state by a pair of droplets.
The head-to-tail alignment of the magnetization vectors of the two
droplets induces a long-range attractive force, whereas the droplets
repel each other in close proximity, resulting in a stable bound state.

We consider spin-$F$ bosonic atoms in free space at zero temperature with the $s$-wave contact interaction and magnetic DDI, which are described by the extended Gross-Pitaevskii equation (eGPE) as
\begin{equation}
\label{eq:gp}
    \begin{aligned}
        i\hbar \frac{\partial\psi_m}{\partial t}
    & = -\frac{\hbar^2}{2M} \nabla^2 \psi_m
  + \frac{4\pi\hbar^2 a_s}{M} \rho \psi_m \\
  & + \frac{32}{3\sqrt{\pi}} \frac{4\pi\hbar^2 a_s^{5/2}}{M}
  \chi(\varepsilon_{\rm dd}) \rho^{3/2} \psi_m \\
  & + g\mu_B (\bm{B} + \bm{B}_{\rm dd}) \cdot \sum_{m^\prime}
  (\bm{S})_{mm^\prime} \psi_{m^\prime},
    \end{aligned}
\end{equation}where $\psi_m(\bm r)$ is the macroscopic wave function
for the magnetic sublevels $m$($= -F, -F+1, ... , F$), $M$ is the mass
of an atom, $a_s$ is the spin-independent $s$-wave scattering length,
and $\rho(\bm r)=\sum_m\rho_m(\bm r)=\sum_m|\psi_m(\bm{r})|^2$ is the
total density normalized as $\int\rho(\bm r)d{\bm r}=N$ with $N$ being
the total number of atoms. We focus on the case in which the DDI
dominates the spin-dependent contact interaction and the spin is
therefore fully polarized everywhere. In this case, the $s$-wave scattering occurs only between atoms fully polarized in the same direction, i.e., the contact interaction can be reduced to the second term on the right-hand side (RHS) of~\cref{eq:gp}, where $a_s$ is the $s$-wave scattering length for a colliding channel with total spin $2F$. For the same reason, we can safely use the Lee-Huang-Yang correction for a single-component dipolar BEC~\cite{Uchino, Yogurt}, which is given by the third term on the RHS of~\cref{eq:gp} with $\chi(\varepsilon_{\rm{dd}})$ being the real part of $\int_0^\pi\sin\theta [1 + \varepsilon_{\rm dd} (3\cos^2\theta - 1)]^{5/2}/2d\theta$. The final term on the RHS of~\cref{eq:gp} represents the linear Zeeman effect arising from the external magnetic field $\bm B$ and the magnetic field $\bm{B}_{\rm dd}$ produced by the DDI as
\begin{equation}
\label{Bdd}
  \bm{B}_{\rm dd}(\bm{r}) = \frac{g \mu_B \mu_0}{4\pi} \int
  \frac{\bm{f}(\bm{r}^\prime) - 3 [\bm{f}(\bm{r}^\prime) \cdot \bm{e}]
    \bm{e}}{|\bm{r} - \bm{r}^\prime|^3}  d\bm{r}^\prime,
\end{equation}
where $g$ is the hyperfine $g$ factor, $\mu_B$ is the Bohr magneton, $\mu_0$ is the magnetic
permeability of vacuum, and $\bm{e}=(\bm r-\bm r^\prime)/|\bm r-\bm r^\prime|$. The spin density is defined by $\bm{f}(\bm{r}) = \sum_{mm^\prime}\psi_m^*(\bm{r})(\bm{S})_{mm^\prime}\psi_{m^\prime}(\bm{r})$, where $\bm{S}$ is the vector of the spin-$F$ matrices. The relative strength of the DDI is characterized by $\varepsilon_{\rm{dd}}=a_{\rm{dd}}/a_s$, where $a_{\rm dd}=\mu_0(g\mu_B)^2M/(12\pi\hbar^2)$. The condition for which the spin-dependent interaction is negligible in~\cref{eq:gp} is $a_s \lesssim a_{\rm dd}$ and $|a_{\rm sd}|\lesssim a_{\rm dd}$, where $a_{\rm sd}$ are the spin-dependent scattering lengths~\cite{Li2024}.

In what follows, we normalize the length, time, density, and magnetic
field by $L_0=a_sN$, $T_0=Ma_s^2N^2/\hbar$, $D_0=1/(a_s^3N^2)$, and
$B_0=\hbar^2/(M a_s^2 N^2 g \mu_B)$, respectively (see Supplemental
Material~\cite{SM} for details of the nondimensionalization). The eGPE
in~\cref{eq:gp} is numerically solved using the pseudospectral method
with spatial step $dx \simeq 10^{-3}$ and time step $dt \simeq
10^{-7}$. To obtain the ground or stationary state,~\cref{eq:gp} is
propagated in imaginary time, in which $i$ on the left-hand side (LHS)
is replaced with $-1$. The following results are qualitatively
independent of $F$; for simplicity, we study the case of $F=1$.

We first present the nonrotating ground state of the self-bound
droplet~\cite{Li2024} for $\bm{B}=0$
in~\subfigsref{fig:panorama}{(a)-1(c)}. The droplet has a torus shape
and the spin vectors circulate along the torus, forming a flux-closure
structure to minimize the magnetostatic energy
[\subfigref{fig:panorama}{(a)}]. We note that this is the ground state
with a negative energy and is robust against external
disturbances. Although $\psi_{\pm1}$ contain vortices [\subfigref{fig:panorama}{(b)}], the droplet has no net orbital angular momentum, $\langle\bm{L}\rangle=-i\int\sum_m\psi_{m}^*(\bm{r})\bm{r}\times\bm{\nabla}\psi_{m}(\bm{r})d\bm{r}=0$, nor net spin angular momentum, $\langle\bm{f}\rangle=\int\bm{f}d\bm{r}=0$. As found from the color of the arrows in~\subfigref{fig:panorama}{(a)}, the spin is fully polarized, $\bm{f}(\bm{r})/\rho(\bm{r})\simeq 1$, and the assumption made in~\cref{eq:gp} is justified.

We next study the rotating droplet state, which is the main purpose of this Letter. We imprint a quantized vortex to the ground-state droplet as
\begin{equation}
    \psi_m(\bm r)=e^{i\ell\theta}\psi_{0,m}(\bm r),\label{eq:phsip}
\end{equation}
where $\ell$ is an integer, $\theta=\arg{(x+iy)}$, and $\psi_{0,m}$ is the ground state in~\subfigsref{fig:panorama}{(a)-1(c)}. Starting from the state in~\cref{eq:phsip}, we perform the imaginary-time evolution and obtain rotating stationary states. The result for $\ell=1$ is shown in~\subfigsref{fig:panorama}{(d)-1(f)}. Similarly to the ground state, the droplet has a torus shape with the circulating spin vectors.
The hole of the torus is larger than that associated with the ground
state because of the centrifugal force.
This state is metastable and robust against external disturbances (see Supplemental Material~\cite{SM}).
The striking feature of the rotating droplet is that it has not only the orbital angular momentum $\langle L_z \rangle\simeq 0.96$ but also the net spin angular momentum $\langle f_z \rangle \simeq 0.04$ in the axial direction of the torus [\subfigref{fig:panorama}{(f)}].
This spontaneous magnetization is due to the imbalance between $m =
\pm 1$ [\subfigref{fig:panorama}{(e)}]; that is, the $m = 1$ and $-1$
states have vorticity $v$ of $0$ and $2$, respectively. As a result,
the $m = 1$ state is more populated to reduce the kinetic energy, giving rise to an increase in $\langle f_z \rangle$.
In this population exchange between different $m$, the total angular
momentum $\langle L_z \rangle + \langle f_z \rangle = 1$ is conserved,
because $m + v = 1$ is satisfied for all~$m$ [\subfigref{fig:panorama}{(e)}], i.e., the total angular momentum per atom is the same for all $m$.
Thus, the spontaneous magnetization arises in the axial direction because of the transfer of the angular momentum from orbital to spin, which resembles the Barnett effect~\cite{Barnett}. 

The nonrotating droplet in~\subfigsref{fig:panorama}{(a)-1(c)} is
achiral, i.e., the parity transformed state~[$\psi_m(\bm{r})
\rightarrow \psi_m(-\bm{r})$] of~\subfigref{fig:panorama}{(a)}, in which the arrows oppositely circulate along the torus, can be superimposed with~\subfigref{fig:panorama}{(a)} itself by the entire $\pi$ rotation of the system around the $x$ axis.
By contrast, for the rotating droplet in~\subfigref{fig:panorama}{(d)}, the parity transformed state has opposite circulation of spin vectors along the torus but its net magnetization is in the same direction ($+z$ direction).
Therefore, the state in~\subfigref{fig:panorama}{(d)} is chiral; namely, the parity transformed state can never be superimposed with the original state by any rotation.
\begin{figure}[t]
    \centering
    \includegraphics[width=1.0\linewidth]{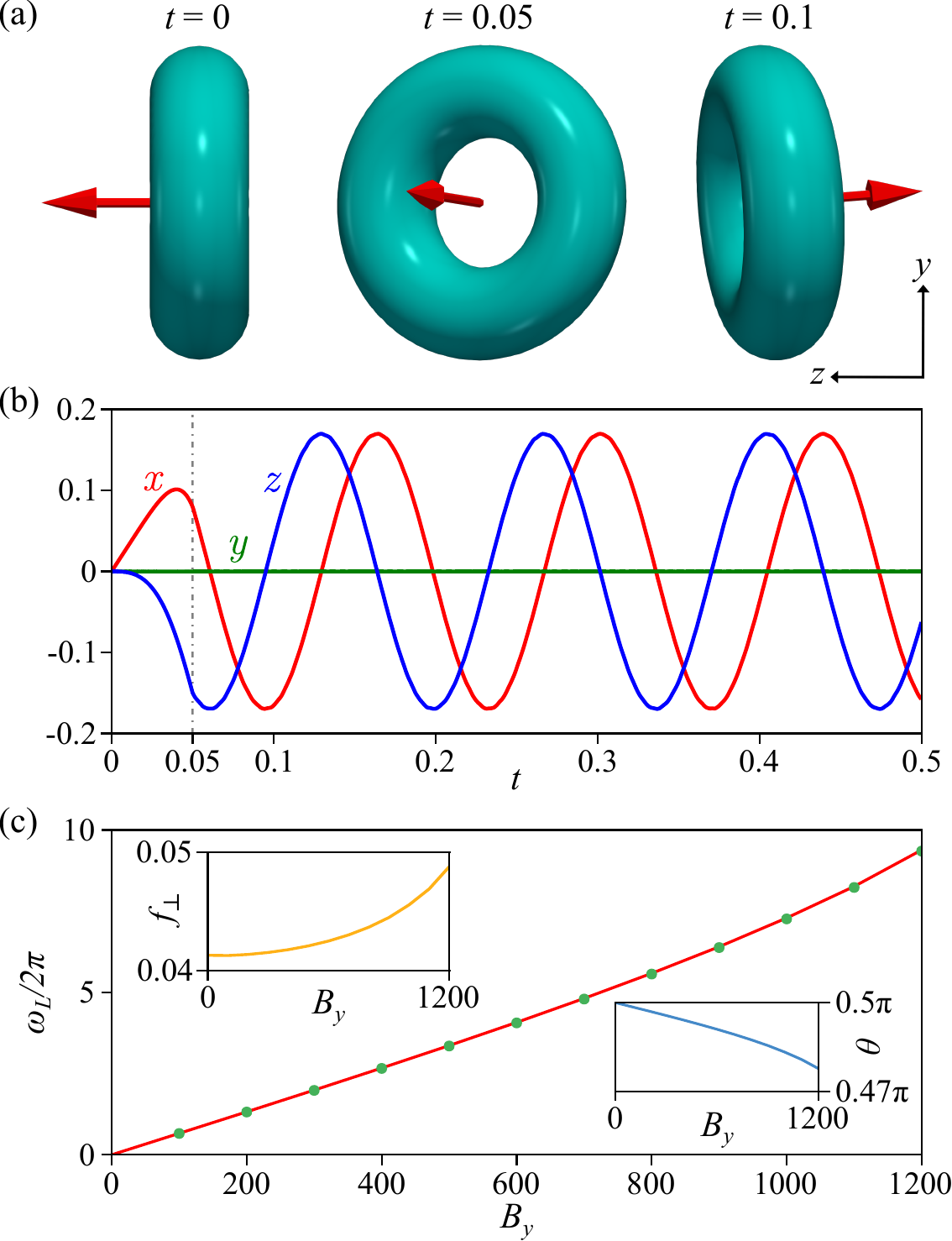}
    \caption{Larmor precession of the vortex droplet in~\subfigsref{fig:panorama}{(d)-1(f)} by an external magnetic field in the $y$ direction. (a) Time evolution of the isodensity surface observed from the $+x$ direction, where $B_y$ is linearly increased from $0$ to $1000$ during the period from $t = 0$ to $0.05$. The arrows indicate the directions of $\braket{\bm L}$. A movie showing the time evolution is provided in the Supplemental Material~\cite{SM}. (b) Time evolution of the $x$, $y$, and $z$ components 
    of~\cref{eq:LP}. (c) Larmor frequency $\omega_{L}$ as a function of $B_y$. The circles in the main panel are obtained by fitting the sinusoidal curves, as in (b). The solid line represents~$\omega_{\rm{L}} = \frac{f_\perp \gamma B_y}{L_\perp + f_\perp}$. The insets plot $f_\perp$ and the angle $\theta$ between $\braket{\bm L}$ and $\bm B$ as functions of $B_y$.
}
    \label{fig:LPcheck}
\end{figure}

We next present two phenomena arising from the spontaneous magnetization of the vortex droplet.
The first phenomenon is the ``mechanical'' Larmor precession in an external magnetic field.~\subfignodotref{fig:LPcheck}{(a)} shows the dynamics of the vortex droplet, where the magnetic field in the $y$ direction is linearly ramped from $B_y = 0$ to $1000$ in the period from $t = 0$ to $0.05$. The torus-shaped droplet starts to rotate around the $y$ axis without changing its shape.
This phenomenon can be described by 
\begin{equation}
    \frac{d\braket{ \bm J}}{d t}=\gamma\braket{\bm f}\times\bm B,
    \label{eq:LP}
\end{equation}
where $\braket{\bm J}=\braket{\bm L}+\braket{\bm f}$ is the total
angular momentum and $\gamma=g\mu_B/\hbar$ is the gyromagnetic
ratio.~\equationnodotref{eq:LP} can be easily derived from the eGPE
in~\cref{eq:gp} by noting that~$d\braket{\bm{J}}/dt=0$ for
$\bm{B}=0$.~\subfignodotref{fig:LPcheck}{(b)} plots the $x$, $y$, and $z$ components 
of~\cref{eq:LP}, which exhibits rotation around the $y$ axis.
Therefore, the mechanism responsible for the droplet rotation is Larmor precession.~The striking difference from the usual Larmor precession in a spinor BEC, in which each atomic spin rotates~\cite{Higbie2005}, is that not only the spin but also the entire atomic cloud rotates as a rigid body.
Such a phenomenon is similar to the precession of a micron-scale single-domain ferromagnet~\cite{Kimball2016}. 

The circles in~\subfigref{fig:LPcheck}{(c)} show the dependence of the Larmor frequency $\omega_{L}$ on the value of $B_y$ after the ramp, where $\omega_{L}$ is obtained by fitting the oscillation in~\subfigref{fig:LPcheck}{(b)} to a sinusoidal curve.
After the applied magnetic field becomes a constant at $t = 0.05$, $L_\perp \equiv \sqrt{\braket{L_x}^2 + \braket{L_z}^2}$ and $f_\perp \equiv \sqrt{\braket{f_x}^2 + \braket{f_z}^2}$ are kept almost constant.
In this approximation,~\cref{eq:LP} can be solved to give~$\omega_{\rm{L}} = \frac{f_\perp \gamma B_y}{L_\perp + f_\perp}$, which is plotted as the line in~\subfigref{fig:LPcheck}{(c)}, in good agreement with the circles.
The function $\omega_L(B_y)$ exhibits a convex curve, since $f_\perp(B_y)$ is an increasing function, as shown in the inset of~\subfigref{fig:LPcheck}{(c)}.
The present mechanism differs substantially from that in
Ref.~\cite{Li2024} (similar rotation of a droplet is shown in Fig.~7
in Ref.~\cite{Li2024}); specifically, the present mechanism is not the Einstein-de Haas effect.
This is easily understood from the fact that a small
transfer of the angular momentum from~$\braket{f_y}$ to~$\braket{L_y}$
cannot rotate the droplet around the $y$ axis because of the gyroscopic effect.

\begin{figure}
    \centering
    \includegraphics[width=1\linewidth]{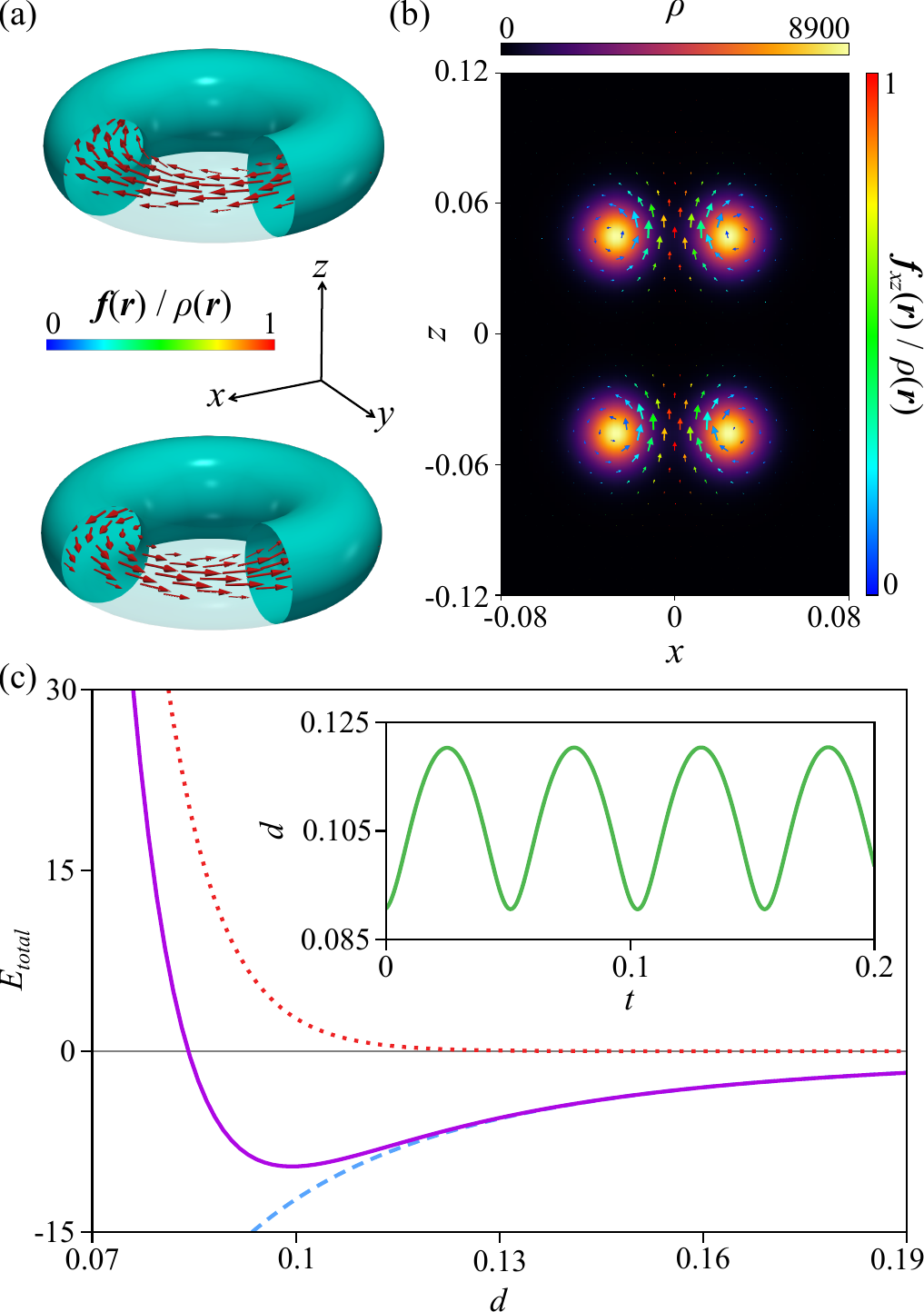}
    \caption{Bound state of two vortex droplets with~$\ell=1$. (a) Isodensity surfaces at half the maximum density for the bound state. (b) Total density distributions and projection of spin vectors on the $y=0$ plane. The color and length of the arrows represent $\bm f_{xz} \left(\bm r\right)/\rho\left(\bm r\right)$ and $|\bm f_{xz}(\bm r)|$, respectively. (c) $E_{\rm total}$ (solid line), $E_{\rm at}$ (dashed line), and $E_{\rm rep}$ (dotted line) as functions of $d$. The inset shows the dynamics of $d$ followed by a slight decrease in $d$ (a movie is provided in Supplemental Material~\cite{SM}). The parameters are the same as those in~\cref{fig:panorama}.}
    \label{fig:bound}
\end{figure}

The second phenomenon arising from the spontaneous magnetization of the vortex droplet is the metastable binding of a pair of droplets. 
We prepare two coaxially aligned vortex droplets that are chirally
different from each other, as shown
in~\subfigpairref{fig:bound}{(a)}{(b)}, where the~$\braket{\bm f}$
and~$\braket{\bm L}$ of both droplets are in the $+z$ direction. As a
result of the imaginary-time evolution, we confirmed that the state in~\subfigref{fig:bound}{(a)} is a metastable bound state of the two droplets.
In fact, the real-time evolution after a small change in the distance $d$ between the droplets exhibits oscillation of $d$, as shown in the inset in~\subfigref{fig:bound}{(c)}, where $d = \int_{z>0} z \rho d\bm{r} / \int_{z>0} \rho d\bm{r} - \int_{z<0} z \rho d\bm{r} / \int_{z<0} \rho d\bm{r}$.
We also confirmed by the real-time evolution that the bound state is also robust against the off-axis displacement of the droplets (see Supplemental Material~\cite{SM}).

The stability arises from the balance between the long-range attractive and short-range repulsive interactions.
The attractive interaction originates from the head-to-tail DDI between the net magnetizations $\braket{f_z}$ of the two droplets, whose energy is estimated to be
\begin{equation}
    E_{\rm at} = -\frac{\mu_0}{4\pi}\frac{2\mu_1\mu_2}{d^3},
    \label{eq:dipole-approx}
\end{equation}
where~$\mu_1=\mu_2=g \mu_B\braket{f_z}$ with~$\braket{f_z}$ being the magnetization of a single droplet.
The repulsive interaction energy is estimated as (see Supplemental Material~\cite{SM} for derivation)
\begin{equation}
    E_{\rm rep}=(1 + \varepsilon_{\rm dd})\frac{4\pi \hbar^2a_s}{M}\int\rho^{(1)}(\bm r)\rho^{(2)}(\bm r)d\bm r,
    \label{eq:Es}·
\end{equation}
where $\rho^{(1)}$ and $\rho^{(2)}$ are the densities of the lower and upper
droplets, respectively.~\subfignodotref{fig:bound}{(c)} shows plots of
$E_{\rm at}$, $E_{\rm rep}$, and $E_{\rm total} = E_{\rm at} + E_{\rm rep}$ as functions of $d$.
There is a minimum in $E_{\rm total}$ at $d\simeq 0.1$, in good agreement with the distance $d$ for the stable bound state.
Thus, the formation of the bound state is due to the long-range attraction between the spontaneous magnetizations of the vortex droplets, in combination with the short-range repulsion arising from the overlap of the two droplets.

The phenomena presented above can be realized experimentally if there is an atomic species that has a stable hyperfine state with~$\varepsilon_{\rm dd} \gtrsim 1$.
The nonrotating droplet in~\subfigsref{fig:panorama}{(a)-1(c)} is the true ground state and can be created by energy relaxation from a suitably prepared initial state~\cite{Li2024}.
The rotating droplet in~\subfigsref{fig:panorama}{(d)-1(f)} can be obtained by the phase imprinting in~\cref{eq:phsip} followed by energy relaxation with total angular-momentum conservation.
The magnetization induced by the Barnett effect can be corroborated by the observation of the mechanical Larmor precession under an external magnetic field~(\cref{fig:LPcheck}).
For concreteness, assuming the $F = 1$ hyperfine state of~$^{151}$Eu atoms with a magnetic moment of $9/2$ Bohr magneton and $\varepsilon_{\rm dd} = 1.2$, the units used above are given by $L_0 = 16.35$ $\mu$m, $T_0 = 0.64$ s, $D_0 = 3.43$ $\mu$m$^{-3}$, and $B_0 = 0.2$ $\mu{\rm G}$.

In conclusion, we have investigated the self-bound quantum droplets with vortices in spinor dipolar BECs. In the presence of a vortex, the mirror-reflection symmetry in the torus-shaped droplet [\subfigsref{fig:panorama}{(a)-1(c)}] is broken, allowing the net magnetization in the direction of the orbital angular momentum [\subfigsref{fig:panorama}{(d)-1(f)}]. The magnetization mechanism is similar to the Barnett effect, in which the orbital angular momentum is transferred to the spin angular momentum.
We demonstrated two phenomena arising from the rotation-induced magnetization of the droplets.
When the external magnetic field is applied to the droplet, it undergoes mechanical Larmor precession (\cref{fig:LPcheck}).
When two chirally different droplets are axially aligned, they form a stable bound state by the attractive interaction between the rotation-induced magnetizations (\cref{fig:bound}).


This work was supported by JSPS KAKENHI Grant~Nos.~JP23K03276 and JP26K00638. SL is supported by the New Energy and Industrial Technology Development Organization (NEDO), Japan (Project No. JPNP20017).

\section{Appendix}
\subsection{Dimensionless quantities}
We normalize the length, time, and wave functions as $\tilde{\bm{r}} =
\bm{r} / L_0$, $\tilde{t} = t \hbar / (M L_0^2)$, and $\tilde{\psi}_m
= L_0^{3/2} N^{-1/2} \psi_m$, respectively, where~$L_0$ is a length unit.
The extended Gross-Pitaevskii equation (eGPE) in Eq.~(1) in the main text then becomes
\begin{equation}
    \begin{aligned}
         i\frac{\partial\tilde{\psi}_m}{\partial \tilde{t}} =& -\frac{1}{2}\tilde{\nabla}^2\tilde{\psi}_m + \frac{4\pi a_s N}{L_0} \tilde\rho \tilde{\psi}_m \\
        &+ \frac{32}{3\sqrt{\pi}} \frac{4\pi\chi a_s^{5/2} N^{3/2}}{L_0^{5/2}} \tilde{\rho}^{3/2} \tilde{\psi}_m \\
        &+ \left( \frac{3N a_{\rm dd}}{F^2 L_0} \tilde{\bm{A}} + \frac{M L_0^2}{\hbar^2} g\mu_B \bm{B} \right) \cdot \sum_{m'} (\bm{S})_{mm'} \tilde{\psi}_{m'},
    \end{aligned}\tag{S1}
\end{equation}
where $\tilde\rho = L_0^3 N^{-1} \rho$ and
\begin{equation}
    \begin{aligned}
        \tilde{\bm{A}} = \int\frac{\tilde{\bm f}-3[\tilde{\bm f}\cdot \bm{e}]\bm{e}}{|\tilde{\bm{r}}-\tilde{\bm{r}}^\prime|^3}d\tilde{\bm{r}}^\prime,
    \end{aligned}\tag{S2}
\end{equation}
with $\tilde{\bm{f}} = L_0^3 N^{-1} \bm{f}$.
In this normalization, the norm satisfies $\sum_m \int |\tilde{\psi}_m|^2 d\tilde{\bm{r}} = 1$.
Setting the length unit to be $L_0 = a_s N$, we obtain
\begin{equation}
    \begin{aligned}
        \label{eq:nondim-gp}
        i\frac{\partial\tilde{\psi}_m}{\partial \tilde{t}}=&-\frac{1}{2}\tilde{\nabla}^2\tilde{\psi}_m + 4\pi\tilde{\rho}\tilde{\psi}_m + \frac{32}{3\sqrt{\pi}}\frac{4\pi\chi}{N}\tilde{\rho}^{3/2}\tilde{\psi}_m \\
        &+\left( \frac{3\varepsilon_{\rm dd}}{F^2} \tilde{\bm{A}} + \tilde{\bm{B}}\right)\cdot \sum_{m'}({\bm S})_{mm^\prime}\tilde{\psi}_{m^\prime},
    \end{aligned}\tag{S3}
\end{equation}
where $\tilde{\bm{B}} =\bm BM a_s^2 N^2 g \mu_B/\hbar^2$. Thus, the independent parameters are $\varepsilon_{\rm dd}$, $N$, and $\tilde{\bm{B}}$.
\subsection{Variational analysis and stability condition}
\begin{figure}[t]
\includegraphics[width=\linewidth]{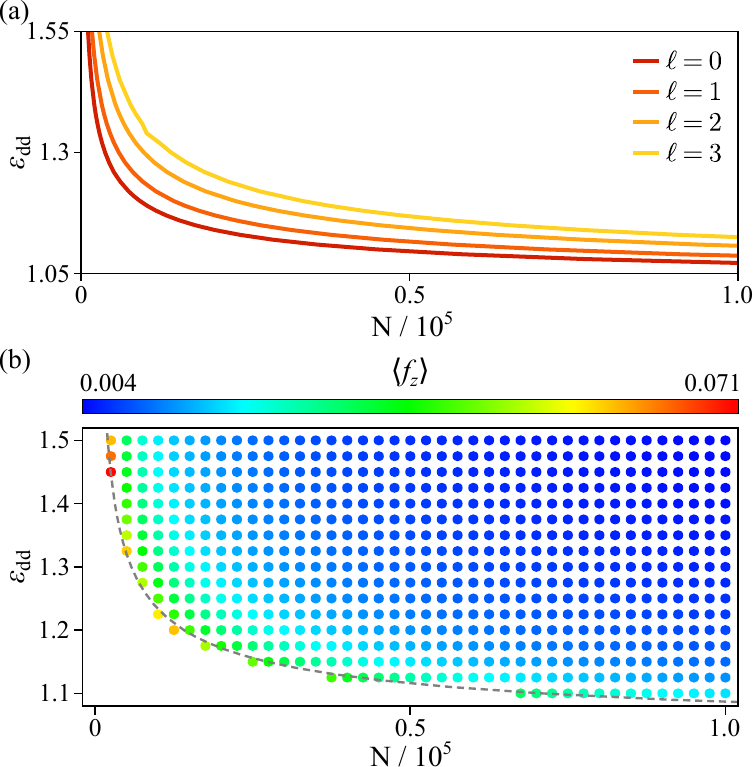} 
\caption{(a) Stability boundary of a vortex droplet with vorticity $\ell$ with respect to $N$ and $\varepsilon_{\rm dd}$ for $F = 1$, obtained by the variational analysis. Stable vortex droplets exist in the regions above the lines. (b) Stability condition (region with circles) for vortex droplet with $\ell = 1$ and $F = 1$, obtained by numerically solving the eGPE.
Color indicates the magnetization $\braket{f_z}$.
The dashed line represents the~$\ell = 1$ line in (a).}
\label{fig:fz}
\end{figure}
We perform the variational analysis to estimate the stability condition of the vortex droplet. We consider a variational wave function as
\begin{equation} 
  \bm{\Psi}_{\rm v}(\bm{r}) = \sqrt{\rho_{\rm v}(r, z)}e^{i\ell\phi}e^{-i {S}_z \phi}
  \bm{\zeta}^{(y)},\label{eq:psiv}\tag{S4}
\end{equation}
where $(r, \phi, z)$ is a position vector in cylindrical coordinates,
the matrix $e^{-iS_z\phi}$ rotates the spin vector around the $z$ axis
by an angle $\phi$, and $\bm{\zeta}^{(y)}$ represents the spin state fully polarized in the $y$ direction with $\sum_m |\bm\zeta_m^{(y)}|^2 = 1$. 
We employ the torus-shaped Gaussian variational function as
\begin{equation}
    \rho_{\rm v}(r, z) = \frac{N}{\pi^{3/2} \sigma_r^{2\lambda + 2} \sigma_z
  \Gamma(\lambda + 1)} r^\lambda
e^{-\frac{r^2}{\sigma_r^2} - \frac{z^2}{\sigma_z^2}},\tag{S5}
\label{eq:rhov}
\end{equation}
where~$\sigma_r > 0$, $\sigma_z > 0$, and $\lambda > 0$ are variational parameters and~$\Gamma$ is the gamma function. The rotation matrix~$e^{-i S_z \phi}$ produces the spin density as~$f_x = -F\rho_{\rm v}\sin\phi$ and~$f_y = F\rho_{\rm v}\cos\phi$, which circulates along the torus, as shown in Fig.~1(d) in the main text.
For the present purpose, we can ignore the small polarization in the
$z$ direction [Fig.~1(f)].
Using the variational wave function, we obtain the kinetic part of the variational energy as
\begin{equation}
\label{eq:Ekin}
E_{\rm kin} = \frac{N \hbar^2}{2M} \left[
  \frac{1}{2\sigma_r^2} \left( 2 + \frac{F+2\ell^2}{\lambda} \right) +
  \frac{1}{2\sigma_z^2} \right].\tag{S6}
\end{equation}
From the form of $E_{\rm kin}$, the variational parameter $\lambda$
that minimizes the variational energy increases with increasing
vorticity $\ell$, and the central hole of the torus becomes larger, which is the manifestation of the centrifugal force.

The contact interaction energy $E_s$, dipole-dipole interaction (DDI) energy $E_{\rm DDI}$, and Lee-Huang-Yang energy $E_{\rm LHY}$ are the same as those for $\ell = 0$, given by~\cite{Li2024}
\begin{align}
    \frac{ME_{\rm ddi}}{N\hbar^2}&=-\varepsilon_{\rm dd}\frac{ME_s}{N\hbar^2}=-\frac{Na_{\rm dd}\Gamma(\lambda+1/2)}{\sqrt{2}\pi\Gamma(\lambda+1)\sigma_r^2\sigma_z},\tag{S7} \\
    \frac{M E_{\rm LHY}}{N \hbar^2}&= \frac{2^{(5\lambda + 17)/2}
    N^{3/2} a_s^{5/2} \lambda \Gamma(5\lambda / 2)
    \chi(\varepsilon_{\rm dd})} {3 \pi^{7/4} 5^{(5\lambda + 3) / 2}
    \Gamma^{5/2}(\lambda + 1) \sigma_r^3 \sigma_z^{3/2}}.\tag{S8}
\end{align}
We minimize the total variational energy $E = E_{\text{kin}} + E_{\text{s}} + E_{\text{ddi}} + E_{\text{LHY}}$ with respect to $\sigma_r$, $\sigma_z$, and $\lambda$ using the Newton-Raphson method. In~\subfigref{fig:fz}{(a)}, we present the curves of critical atom number $N$ estimated for $F = 1$ and $\ell=0,1,2,$ and $3$. In the regions under the curves, the atomic cloud fails to remain self-confined and inevitably expands.

We also investigate the stability condition for~$\ell = 1$ using the
imaginary-time propagation of the eGPE [\subfigref{fig:fz}{(b)}]. The
stability boundary obtained by the eGPE agrees well with that obtained
by the variational analysis.
The magnetization $\braket{f_z}$ becomes larger near the stability
boundary because the DDI effect is relatively small and deviation from
the flux closure structure becomes easier.

\subsection{Interaction between two vortex droplets}
We derive Eq.~(6) in the main text, which expresses the repulsive
interaction between the two vortex droplets aligned as in Fig.~3(a).
Since the long-range attractive interaction arising from the net
magnetizations $\langle f_z \rangle$ of the two droplets has already
been taken into account as $E_{\rm at}$ in Eq.~(5), we ignore the
small magnetization in the $z$ direction and assume that the spins are
fully magnetized in the $x$-$y$ directions, circulating along the
tori.
We write the wave function of the two vortex droplets as
\begin{equation}
  \psi_m^{\rm (tot)}(\bm{r})
  = \psi_m^{(1)}(\bm{r}) + \psi_m^{(2)}(\bm{r}),\tag{S9}
\end{equation}
where the superscripts $(1)$ and $(2)$ distinguish the two vortex
droplets.
Let us consider a situation in which the two droplets come close to
and overlap with each other.
Since the fully-polarized spin density circulating along torus 1 is opposite to
that in torus 2, their spin states at $\bm{r}$ are orthogonal to
each other as
\begin{equation}
    \begin{aligned}
        \sum_m \psi_m^{(1)*}(\bm{r}) \psi_m^{(2)}(\bm{r}) & = 0,\\
        \sum_{m, m'} \psi_m^{(1)*}(\bm{r}) (\bm{S})_{mm'}
        \psi_{m'}^{(2)}(\bm{r}) & = 0.
    \end{aligned}\tag{S10}
\end{equation}
The total density is therefore written as
\begin{equation}
    \begin{aligned}
          \rho^{\rm (tot)}(\bm{r}) & = \sum_m |\psi_m^{\rm (tot)}(\bm{r})|^2
  \nonumber \\
  & = \sum_m |\psi_m^{(1)}(\bm{r})|^2 + \sum_m |\psi_m^{(2)}(\bm{r})|^2
  \nonumber \\
  & \equiv \rho^{(1)}(\bm{r}) + \rho^{(2)}(\bm{r}).
  \label{rho}
    \end{aligned}\tag{S11}
\end{equation}
Similarly, the total spin density has the form
\begin{equation}
    \begin{aligned}
          \bm{f}^{\rm (tot)}(\bm{r}) = & \sum_{m, m'}
  \psi_m^{\rm (tot)*}(\bm{r}) (\bm{S})_{mm'}
  \psi_{m'}^{\rm (tot)}(\bm{r})
  \nonumber \\
  = & \sum_{m, m'} \psi_m^{(1)*}(\bm{r}) (\bm{S})_{mm'}
  \psi_{m'}^{(1)}(\bm{r})
  \nonumber \\
  & + \sum_{m, m'} \psi_m^{(2)*}(\bm{r}) (\bm{S})_{mm'}
  \psi_{m'}^{(2)}(\bm{r})
  \nonumber \\
  \equiv & \bm{f}^{(1)}(\bm{r}) + \bm{f}^{(2)}(\bm{r}),
  \label{f}
    \end{aligned}\tag{S12}
\end{equation}
where $\bm{f}^{(1)}(\bm{r})$ and $\bm{f}^{(2)}(\bm{r})$ circulate
along the tori in the opposite directions, i.e.,
$\bm{f}^{(1)}(\bm{r}) \propto -\bm{f}^{(2)}(\bm{r})$.
Substituting Eq.~(\ref{rho}) into the $s$-wave contact interaction
energy,
\begin{equation} \label{Estot}
  E_s^{\rm (tot)} = \frac{2\pi\hbar^2 a_s}{M} \int
  \left[ \rho^{\rm (tot)}(\bm{r}) \right]^2 d\bm{r},\tag{S13}
\end{equation}
we find that the contact interaction energy between the two droplets
corresponds to
\begin{equation} \label{Econtact}
  E_s^{(12)} = \frac{4\pi\hbar^2 a_s}{M} \int
  \rho^{(1)}(\bm{r}) \rho^{(2)}(\bm{r}) d\bm{r}.\tag{S14}
\end{equation}
Substituting Eq.~(\ref{f}) into the DDI energy,
\begin{widetext}
\begin{equation}
  E_{\rm ddi}^{\rm (tot)} = \frac{\mu_0 (g\mu_B)^2}{8\pi}
  \int d\bm{r} d\bm{r}'
  \frac{\bm{f}^{\rm (tot)}(\bm{r}) \cdot \bm{f}^{\rm (tot)}(\bm{r}')
    - 3 [ \bm{f}^{\rm (tot)}(\bm{r}) \cdot \bm{e} ]
    [ \bm{f}^{\rm (tot)}(\bm{r}') \cdot \bm{e} ]}{|\bm{r} -
    \bm{r}'|^3},\tag{S15}
\end{equation}
gives the DDI energy between the two droplets as
\begin{equation}
  E_{\rm ddi}^{(12)} = \frac{\mu_0 (g\mu_B)^2}{4\pi}
  \int d\bm{r} d\bm{r}'
  \frac{\bm{f}^{(1)}(\bm{r}) \cdot \bm{f}^{(2)}(\bm{r}')
    - 3 [ \bm{f}^{(1)}(\bm{r}) \cdot \bm{e} ]
    [ \bm{f}^{(2)}(\bm{r}') \cdot \bm{e} ]}{|\bm{r} - \bm{r}'|^3}.\tag{S16}
\end{equation}
\end{widetext}
For the form of the wave function in~\cref{eq:psiv}, $E_{\rm ddi}^{(12)}$
becomes~\cite{Li2024}
\begin{equation} \label{Eddi}
  E_{\rm ddi}^{(12)} = \frac{4\pi\hbar^2 a_{\rm dd}}{M}
  \int \rho^{(1)}(\bm{r}) \rho^{(2)}(\bm{r}) d\bm{r}.\tag{S17}
\end{equation}
Equations~(\ref{Econtact}) and (\ref{Eddi}) give Eq.~(6) in the main
text.

\subsection{Stability of vortex droplets}
\begin{figure}[h]
\includegraphics[width=\linewidth]{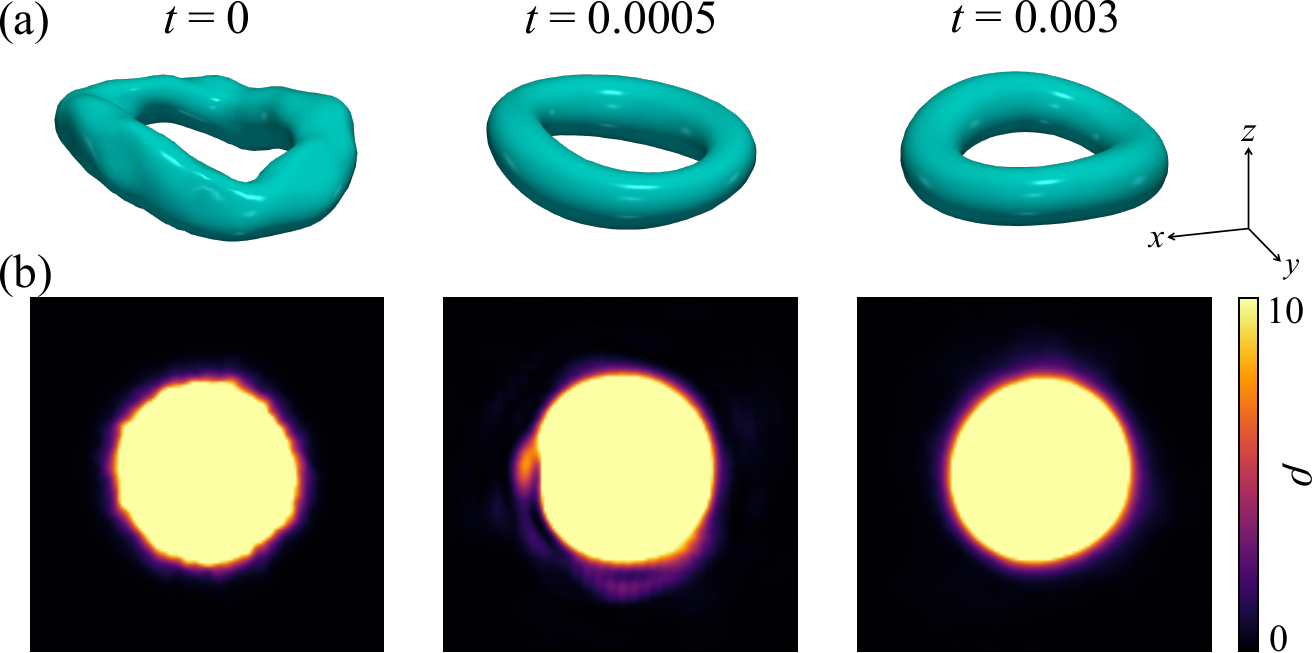}
\caption{Stability of the vortex droplet against external disturbances. (a) Isodensity surface at half the maximum density. (b) Density profiles on the $z=0$ plane.
The central hole is invisible because of the color range. The size of each panel is $0.3\times0.3$. A movie showing the time evolution is provided in the Supplemental Material~\cite{SM}. The parameters are the same as those in~Figs.~1(d)-1(f) in the main text. Low-pass-filtered random noise with a cutoff wave number $2\pi / 0.01$ is added to the initial state.}
\label{fig:stable}
\end{figure}
We here study the stability of the vortex droplets against external
disturbances. We prepare the vortex droplet in~Figs.~1(d)-1(f) in the
main text and apply low-pass-filtered random noise with a cutoff wave
number of $2\pi/0.01$ [\subfigref{fig:stable}{(a)}]. First, a small
number of atoms are emitted from the vortex droplet, as shown
in~\subfigref{fig:stable}{(b)} ($t=0.0005$); the torus-shaped vortex
droplet is thereafter maintained.
A small excitation with a period of $\simeq 0.003$ remains; the energy
of this excitation is smaller than the chemical potential~$|\mu|\simeq 2\pi/0.0002$. 
Thus, we have confirmed that the vortex droplet shown in Fig. 1 is rather robust against external disturbances. We also confirmed the stability of vortex droplets with other vorticity $\ell$ (data not shown).

\begin{figure}[h]
\includegraphics[width=\linewidth]{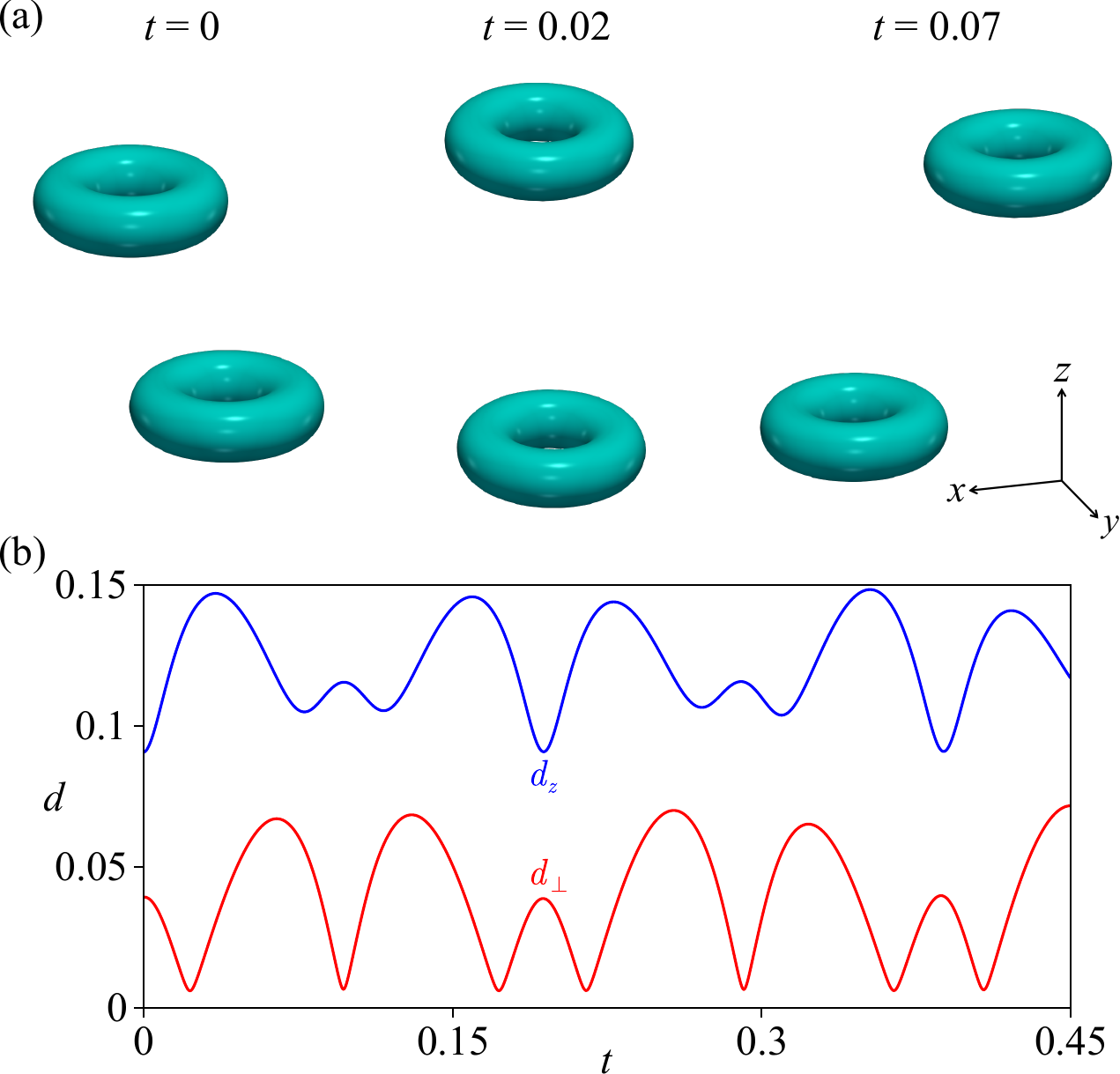}
\caption{Dynamics of bound droplets with initial displacement $d_x=
  0.04$, $d_y=0$, and $d_z=0.09$. (a) Isodensity surfaces at half the
  maximum density. (b) Time evolution of distance between two
  droplets, where $d_z$ and $d_\perp$ represent the distances in the
  $z$ and $x$-$y$ directions, respectively. A movie of the dynamics is
  provided in the Supplemental Material~\cite{SM}. The parameters are the same as those in~Fig.~3 in the main text.}
\label{fig:shift}
\end{figure}

We also investigate the off-axis stability of the bound state in~Fig.~3.
Initially, we displace the position of the one droplet by~$d_x$,~$d_y$, and~$d_z$ in the~$x$, $y$, and $z$ directions, respectively, and perform the real-time evolution.
As shown in~\cref{fig:shift}, the relative position of the vortex droplets oscillate not only in the $z$ direction but also in the $x$-$y$ direction, indicating the robustness of the bound state.
~\\

\end{document}